\documentclass[preprint,authoryear,12pt,twocolumn]{elsarticle}

\usepackage{epsfig}
\usepackage{amssymb}
\usepackage{aas_macros}

\usepackage[ps2pdf,%
a4paper=true,%
breaklinks=true,%
colorlinks=true,%
pdfauthor={First Author et al.},%
pdftitle={Template for manuscripts in Advances in Space Research}%
]{hyperref}

\journal{Proceedings of an International Workshop `Wolf-Rayet Stars'} 
 
\biboptions{authoryear}

\begin{document}

\begin{frontmatter}



\title{X-ray emission from single Wolf-Rayet stars}


\author{Lidia M. Oskinova}
\address{Institute for Physics and Astronomy, University of Potsdam,
14476 Potsdam, Germany}
\ead{lida@astro.physik.uni-potsdam.de}



%
\end{frontmatter}

\noindent
{\em Abstract}: 
In this review I briefly summarize our knowledge of the X-ray emission from 
single WN, WC, and WO stars. 
These stars have relatively modest X-ray luminosities, typically not 
exceeding 1\,$L_\odot$. The analysis of X-ray spectra usually reveals 
thermal plasma with temperatures reaching a few $\times 10$\,MK.  X-ray 
variability is detected in some WN stars. At present we don't fully 
understand how X-ray radiation in produced in WR stars, albeit there are 
some promising research avenues, such as the presence of CIRs in the winds 
of some stars. To fully understand WR stars we need to unravel mechanisms of 
X-ray production in their winds.  

\section{X-rays from WR stars: history}

The discovery of X-ray emission from WR-stars was among the first results 
achieved by the {\em Einstein} X-ray observatory 
\citep{Seward1979}.
This confirmed the theoretical prediction that WR stars, 
specifically WR binaries, are X-ray sources 
\citep{Cherep1976}. \citet{Pollock1987} summarized the
{\em Einstein} observations of WR stars and  
outlined their key X-ray properties. From the analysis of the observations it 
was concluded that ({\it i}) the X-ray brightest WR stars are often massive 
binaries; ({\it ii}) single WR stars are usually faint; ({\it iii}) the WN 
stars are on average more X-ray bright than WC stars. These conclusions remains 
valid today.  

The {\em Rosat} X-ray telescope 
performed an all sky survey. Besides increasing the number of detected 
WR stars, it also obtained X-ray spectra and light curves for some key targets  
\citep[e.g.][]{Willis1996}. A preliminary X-ray catalog of WR stars   
was presented by \citet{Pollock1995}.  For single WR stars no correlation 
between X-ray luminosity and  their bolometric luminosity or 
wind momentum was found \citep{Wes1996}. This result was later on 
confirmed by the detailed studies \citep{Ignace1999,Ignace2000}.

The ASCA X-ray telescope 
observed four WR stars in great details, yielding  X-ray spectra and 
light-curves. Monitoring of massive binaries revealed the unambiguous 
evidence that the X-rays are produced 
in the wind-wind collision zone \citep[e.g.][]{Stevens1996}. 
Other X-ray observatories, such as EXOSAT, also 
contributed to the studies of X-rays from WR-type stars \citep[e.g.][]{Wil1987}.

Today, a fleet of X-ray telescopes operating in space routinely observes 
WR stars. Modern X-ray telescopes have broad pass-bands and significantly 
improved sensitivity, spectral and spatial resolution. Among these 
telescopes, {\em Chandra} and {\em XMM-Newton} have unprecedented capabilities
allowing in depth studies of the formation and propagation of X-rays 
in WR stars. The future also looks bright in X-rays with new missions such as 
Spektrum-R\"ontgen-Gamma (SRG), Astro-H, and Athena being under 
development.

\section{General properties of X-rays from WR stars}

\begin{figure}[t]
\begin{center}
\includegraphics[width=\columnwidth]{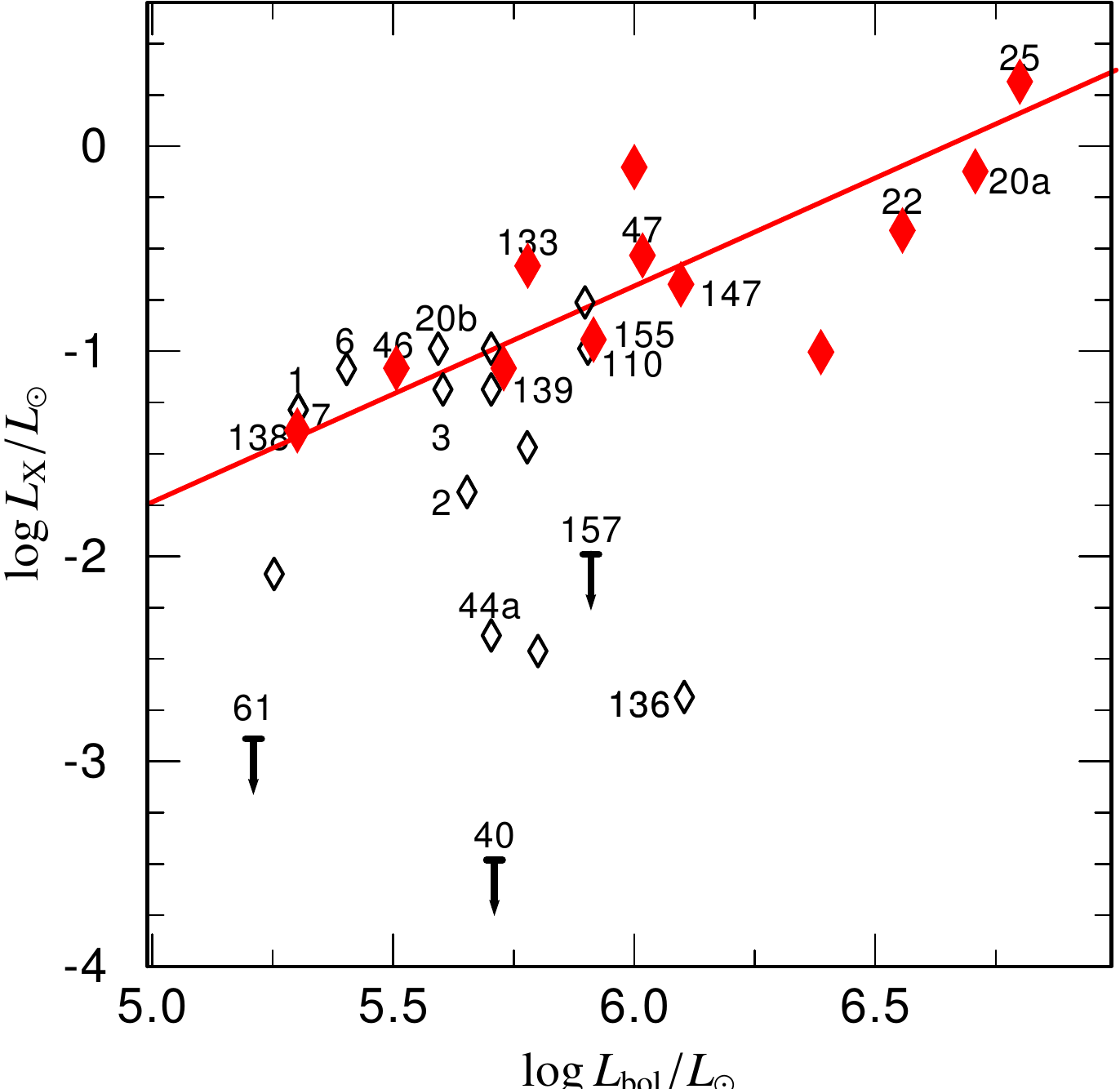}
\caption{$L_{\rm X}$ vs. $L_{\rm bol}$ for WN stars. Black empty 
diamonds denote single stars, while red filled diamonds denote  
binaries \citep{osk2005,Gosset2005,Skin2012,Zhekov2012}.  
The straight line shows the 
correlation for  O-star  binaries 
\citep{osk2005}.
\label{fig:osklxb}}
\end{center}
\end{figure}

{\em Luminostiy.} The lack of
correlation between X-ray and bolometric luminosity of WR stars 
is in a strong contrast to O-type stars. The X-ray luminosities of the 
latter are $L_{\rm X}\approx 10^{-7}L_{\rm bol}$ \citep{Pallavicini1981}. This 
correlation holds  for single as well as for binary O stars 
\citep{osk2005,naze2009}. The origin of this correlation in O stars 
is not yet understood and might be related to the properties of shocks or 
to  magnetic effects \citep{osk2011,owocki2013}.  
As can be seen in Fig.\,\ref{fig:osklxb}, the X-ray luminosities of WN-type 
stars are diverse. While WN binaries seem to show a trend 
similar to the O+OB binaries, the putatively single WN stars  have X-ray 
luminosities that differ by orders of magnitude. Overall, the X-ray 
luminosities 
of WN stars do not significantly exceed $\sim 10^{34}$\,erg\,s$^{-1}$ with 
binaries being on average more X-ray bright.

{\em Variability.} The X-ray monitoring of WR stars revealed that single as 
well as binary stars are X-ray variable. The X-ray variability of binaries is 
on 
the orbital period time scale and is well documented, studied, and understood 
\citep[among most recent papers, e.g.][]{Pandey2014,Lomax2015,Zhekov2015}. The 
sample of single WR stars that were monitored in X-rays is significantly 
smaller. Data of high fidelity exist so far only for WR6 \citep{Ignace2013}.
This WN4 star shows X-ray variability on the level of 20\%\ and with a 
characteristic period similar to the 3.766 day period well 
known from the optical \citep[e.g.][]{Morel1997}, but is not in phase. This is
quite similar to the X-ray variability observed in single O-type stars 
\citep[e.g.][]{osk2001,naze2013}. The origin of the X-ray variability is likely 
related to the presence  of corrotating interaction regions 
(CIRs) in the stellar winds \citep{Chene2011,massa2014}.

{\em Temperature.} Already {\em Einstein} and {\em Rosat} observations revealed 
that X-ray spectra of WR stars can be described as thermal. In binaries, the 
ionization balance could be out of equilibrium and shocks could be 
collisionless \citep{Pollock2005,Zhekov2007}. In single stars,  the 
X-ray spectra were, so far, well reproduced by multy-temperature thermal 
plasmas in collisional equilibrium. Temperatures  between 1\,MK up to 50\,MK 
are found from spectral analyses, with the emission measures of cooler
plasma components being larger than that of hotter ones (the differential 
emission measure declines with temperature) 
\citep{Skina2002,Skinb2002,Skin2010,Ignace2003,osk2012}. 

\section{X-ray production in WR stars}

The present day sophisticated models are capable to describe and explain 
the X-ray emission from WR binaries 
\citep[e.g.][]{Stev1992,Pittard2006,Zhekov2012,Parkin2013,Rauw2015}. Albeit 
the exact physics is not yet fully understood, it is beyond reasonable doubt 
that the bulk of X-rays in the majority of WR binaries is produced in the 
wind-wind collisions zone.  

The situation is very different for single WR stars. There are neither 
theories nor quantitative models explaining the production of X-rays in 
the winds of these stars, albeit there is no shortage of scenarios. 
Among them:

\smallskip\noindent
{\it (i)} X-ray emission  in some ``normal'' WR stars could be due to 
the wind accretion on a compact object \citep{WL1986}. Modern observations do 
not support this scenario. The observed X-ray properties of WR stars differ 
significantly from the X-ray emission of high-mass X-ray binaries. 
Only a few WR stars with relativistic companions are known, and their X-ray 
properties  do not resemble those of single WR stars \citep{Barnard2008}. 

\smallskip\noindent
{\it (ii)} The binarity idea  was further examined by \citet{Skina2002} who 
suggested that a WR wind shocking onto a otherwise unseen close stellar 
companion could be responsible for the X-ray emission in at least some 
apparently single WR stars.
There are a number of arguments questioning this scenario. For instance, 
new deep X-ray observations show no spectral signatures characteristic 
for low-mass stars. The X-ray variability patterns also do not support 
this scenario \citep{osk2012,Ignace2013}.  

\smallskip\noindent
{\it (iii)} Particle acceleration in shocks. According to this scenario, a
population of relativistic electrons accelerated in the wind shocks is 
responsible for the generation of X- and $\gamma$-rays via inverse Compton 
scattering
\citep[e.g.][]{Pollockb1987}. It seems that this mechanism may indeed operate 
in colliding wind binaries, however no observational evidence in its support  
was so far found in single WR stars.  

\smallskip\noindent
{\it (iv)} If a strong dipolar magnetic field is present, it may confine 
stellar winds and lead to strong shocks at the magnetic equator  
\citep{Babel1997}. Albeit this mechanism is sometimes considered in the 
literature to explain the X-ray emission from WR stars,  no strong 
magnetic fields capable to confine powerful WR-winds were detected so far
\citep[][Hubrig et al. 2016, in press]{Chev2014}. Nevertheless, even if the
wind confinement model does not explain X-rays from WR stars, 
the role of magnetism for the X-ray generation cannot be ruled out. E.g.\ a 
strong magnetic field was suspected in a few peculiar WR stars, such as WR2 
\citep{Shenar2014}, but their X-ray properties are not outstanding 
\citep{osk2005,Skin2010}. 
 
\smallskip\noindent
{\it (v)} Shock heating due to line-driven wind instabilities (LDI) is often 
invoked to explain X-ray emission from OB-type stars \citep{feld1997}. 
\citet{GO1995}
showed that multiple scattering in dense WR winds reduces this 
instability. However, the
residual instability may still lead to  clumping of WR winds. These 
authors did not address X-ray emission from WR stars, i.e.\ they did not show 
that the instability in WR winds is sufficient to drive strong shocks where the 
plasma can be heated to a few MK. Yet, it is tempting to draw parallels between 
WR and OB supergiant X-ray production mechanisms (Gayley 2016, submitted). 

\smallskip\noindent
{\it (vi)} \citet{Gayley2012} 
hypothesized that the fast wind may ram into the slow moving clumps, 
whose signatures are  observed as moving bumps on top of lines observed 
in optical WR spectra \citep[][]{Lepine1999}. Plasma could be heated in the 
resulting shocks. This is an interesting scenario that shall be tested 
by detailed modeling and observations. 

\smallskip\noindent
{\it (vii)} Observations confirm that corotating interaction regions (CIRs) are 
present in WR winds \citep{Chenea2011}. Their origin is not fully understood 
but may be related to the presence of surface magnetic fields \citep{Mich2014}.
\citet{Mullan1984,Chene2011} and \citet{Ignace2013} invoked CIRs to explain 
the X-ray generation in WR winds. The detailed hydrodynamic models of the CIRs 
in OB winds \citep{Cranmer1996,Lobel2008} are so far isothermal, so it remains 
to be seen  whether CIRs can be the origin of X-rays from OB star winds. For 
the WR winds, to our knowledge, no hydrodynamic modeling of CIRs 
was performed so far. Yet it seems that the CIR scenario is the most 
promising one to explain the X-ray emission it least in some rotating single 
WR stars.    

\section{Propagation of X-rays}

\begin{figure}[t]
\begin{center}
\includegraphics[width=\columnwidth]{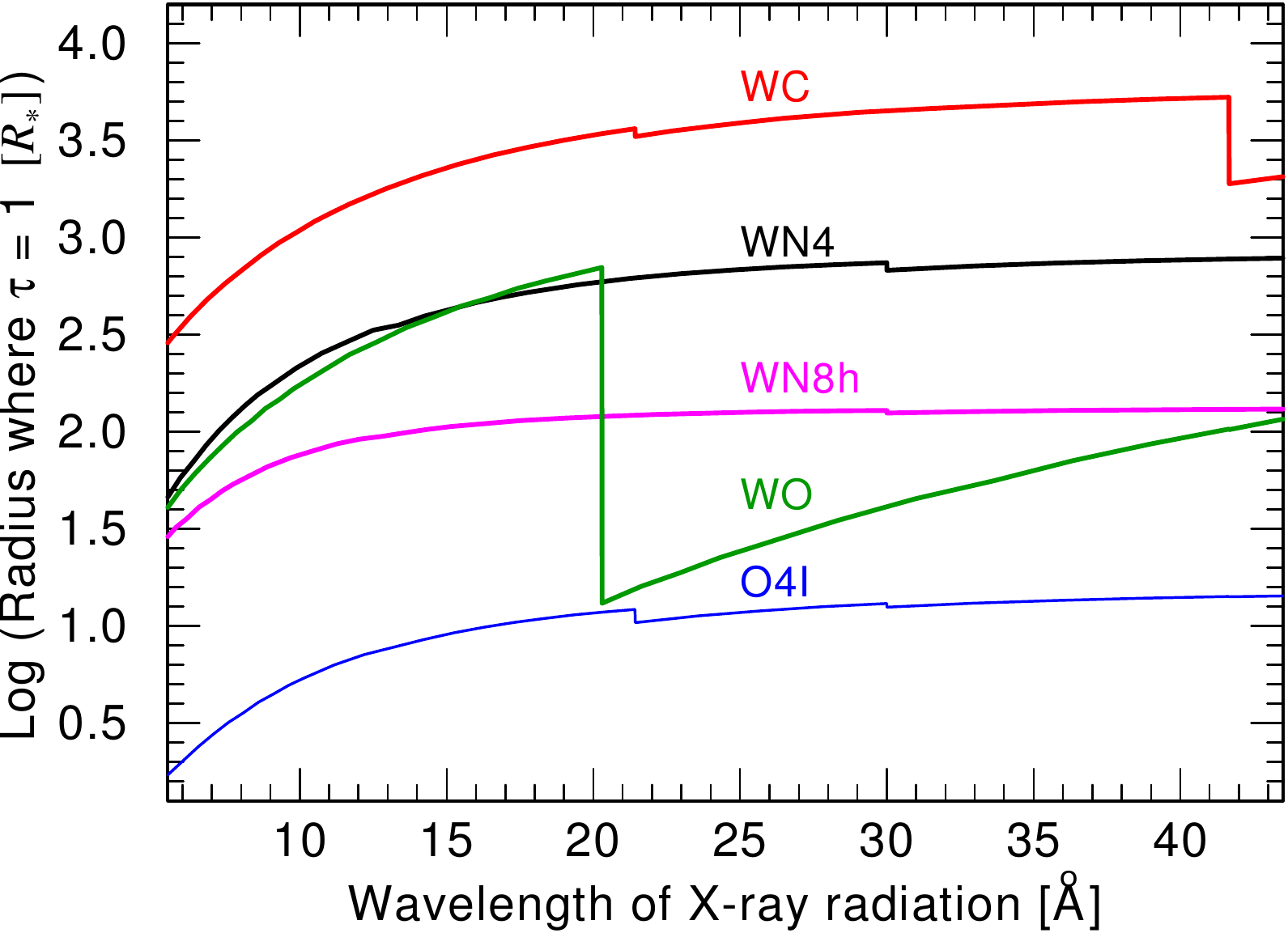}
\caption{Radius where the continuum optical depth 
reaches unity, as predicted by PoWR models for the "cool" wind component 
in different WR stars and an O-supergiant, in dependence 
on wavelength.
\label{fig:osktau1}}
\end{center}
\end{figure}

The combination of large column densities and high metal abundance makes WR 
winds very opaque for X-rays \citep[e.g.][]{Ignace1999}. 
Figure\,\ref{fig:osktau1} demonstrates that WR winds may remain optically thick 
for X-rays up to  a few $\times 1000\,R_\ast$. Wind clumping allows X-rays 
to emerge from somewhat deeper wind layers due to the porosity effects  
\citep{Shaviv2000,feld2003,osk2004}.  

The X-ray field can be included in non-LTE stellar atmosphere models, 
such a PoWR \citep{Baum1992} allowing to consistently solve the 
X-ray transfer in WR winds. This has important implications for 
applying X-ray spectral diagnostics, such as the widely used ratios of 
forbidden to intercombination lines in the X-ray spectra of He-like ions 
\citep[][Huenemoerder et al. 2016, in press]{Leu2006,WC2007}. 

\section{X-rays from  WNE stars}

\begin{figure}[t]
\begin{center}
\includegraphics[width=\columnwidth]{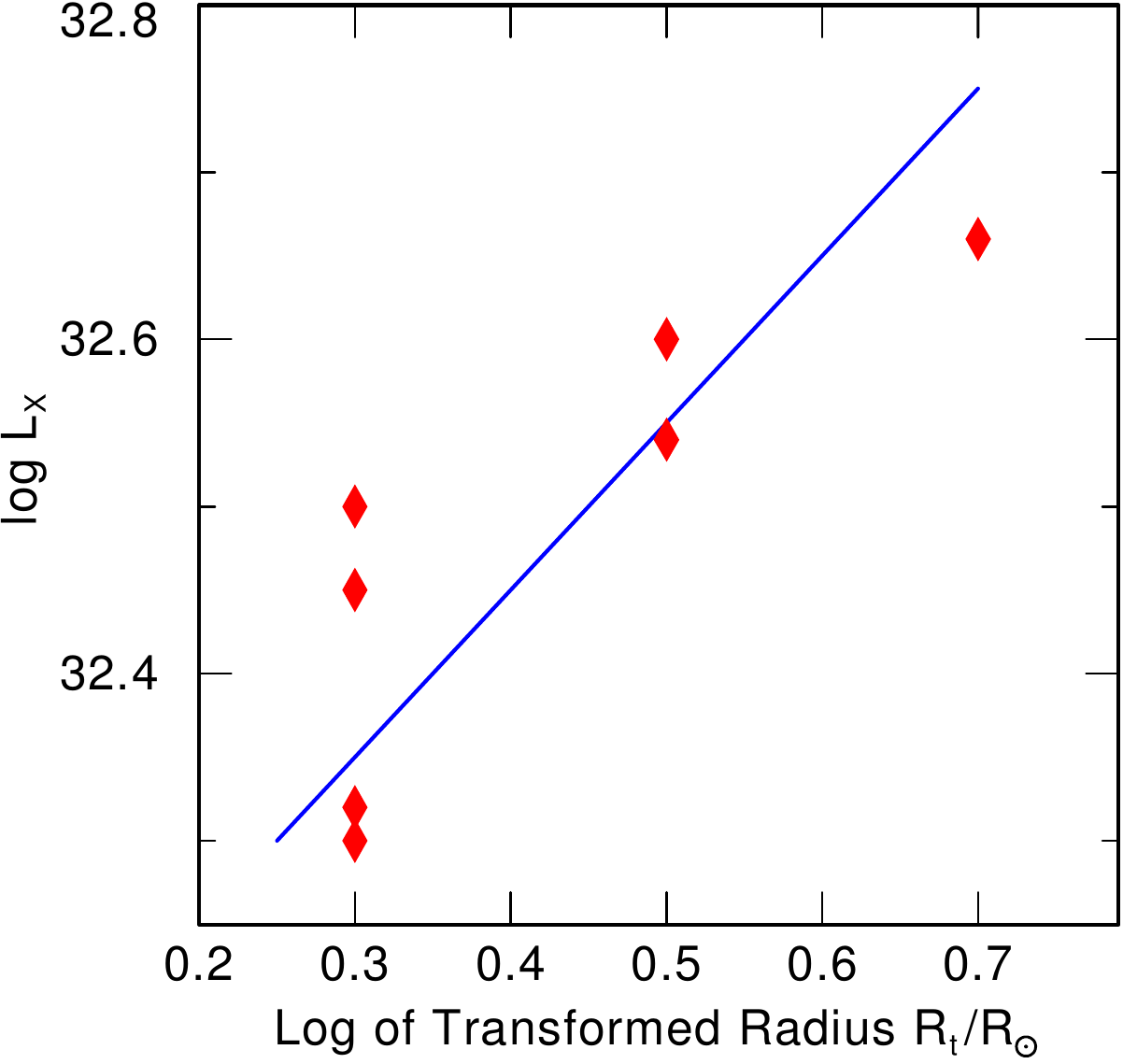}
\caption{X-ray luminosity versus transformed radius $R_{\rm t}$ for a sample of 
WNE stars (red diamonds. The values are from 
\citet{Skina2002,Skinb2002,Skin2010,Ignace2003,osk2005,Hamann2006}. The 
straight blue line shows a linear regression fit. 
\label{fig:oskrt}}
\end{center}
\end{figure}

Hydrogen free early-type WN stars (WNEs) occupy a distinct region in the HRD 
and have rather similar stellar parameters \citep{Hamann2006}. It appears that 
the X-ray properties of WNEs are also similar. The rate of X-ray detections 
of WNE stars is quite high -- all stars that were observed so far with pointing 
observations were detected. Those WNE stars have similar X-ray 
luminosities ($L_{\rm X} \approx 2\,...\,6 \times 10^{32}$\,erg\,s$^{-1}$. 
Their X-ray   temperatures are also similar, with the  emission measure 
weighted average temperature of $\left<T\right>\approx 5$\,MK. The parameter 
$R_{\rm t}$ \citep[see its definition and discussion 
in][or in these Proceedings]{Hamann2006} is used to characterize the emission 
measure of WR winds. Figure\,\ref{fig:oskrt} shows an interesting trend between 
the X-ray luminosity of WNE stars and their $R_{\rm t}$.  The presence of such 
trend implies that the generations of X-rays is an intrinsic property of 
stellar winds.

A high-resolution X-ray spectrum is available, so far, for only one  
single WR star, WR6 (WN4) (see also D. Huenemoerder, 
these proceedings). Its spectra were obtained by both {\em XMM-Newton}
\citep{osk2012}
and {\em Chandra} (Huenemoerder et al. 2016, submitted). The abundances in 
X-ray emitting
plasma are consistent with WNE abundances, e.g.\ no oxygen lines are 
detectable. This provides a strong argument against a non-degenerate 
binary component scenario.  The X-ray lines are broad, blue-shifted,  and have
a characteristic shape as expected when the radiation forms in outer wind
and suffers strong continuum absorption 
\citep{Mac1991,Ignace2001}. The analysis of line ratios in He-like ions
confirms that the radiation is formed far out in the wind. Even the hottest 
plasma is formed at more than tens stellar radii.  
The spectral analysis reveals that X-ray emitting gas must be present even at 
$>1000\,R_\ast$. Such very extended spatial distribution is very difficult to 
reconcile with the predictions of the LDI models 
developed for O stars. 
Interestingly, there are indications that the fluorescent Fe\,K$\alpha$ line 
is present in the X-ray spectrum. This indicates that dense and cool matter 
coexists with the hot plasma.   

\section{X-rays from WNL stars}

Late type WN stars (WNLs) have on average higher bolometric 
luminosities and slower winds than WNE stars \citep[e.g.][]{Sanders1985}.  Some 
WNLs may be  descendants of more massive progenitors than WNEs, 
while others may be on earlier evolutionary stage  \citep{Crow1995,Hamann2006}. 
 \citet{Gosset2005} 
obtained deep {\em XMM-Newton} observations of WR40 (WN8h) and its nebula, but 
neither was detected in X-rays. The wind of WR40  is more transparent for 
 X-rays than the wind of WR1 (WN4)  (see corresponding curves in 
Fig.\,\ref{fig:osktau1}), yet the later is  $>100$ times more X-ray luminous 
than the former. On the other hand, the WN9h star   
WR79a was detected with $L_{\rm X}\approx 10^{32}$\,erg\,s$^{-1}$ 
\citep{osk2005}. Its low resolution X-ray spectrum bears similarities with 
spectra of WNEs, i.e.\ a relatively high-temperature component 
($T\sim 30$\,MK) is present \citep{Skin2012}. Other WNLs  well 
observed in X-rays are WR78 (WR7h) \citep[$L_{\rm X}\approx 
10^{31}$\,erg\,s$^{-1}$][note that \citet{Skin2012} report an order of 
magnitude higher luminosity for this object]{Pollock1995,Wes1996,osk2005} and 
WR136 (WN6(h)) \citep[$L_{\rm X}\approx 
10^{31}$\,erg\,s$^{-1}$,][]{Wrigge1994,Pollock1995,Ignace2000}. 
Thus, while some WNLs are not detected in X-rays, others appear to be 
relatively X-ray luminous. 

\section{X-rays from WC and WO stars}

\citet{Rauwa2015} 
discovered X-rays from the WC4 star WR144, making it the first single WC 
star detected in X-rays. Its low X-ray luminosity ($L_{\rm 
X}\approx 
10^{30}$\,erg\,s$^{-1}$)  corroborates   \citet{osk2003}'s 
conclusion that WC stars are intrinsically X-ray faint, likely because of the 
high opacity of their stellar winds (see Fig.\,\ref{fig:osktau1}). 
On the other hand, known WC binaries are  bright X-ray sources 
\citep[e.g.][]{Schild2004,Pollock2005,Zhekov2011}.
Hence, X-rays provide a convenient indication of binarity:  X-ray 
bright WC stars should be colliding wind binaries. This makes X-rays 
observations a very useful diagnostic tool to search for binaries among WC 
populations \citep[][]{Clark2008, osk2008, Hyodo2008, Mau2011,nebot2015}.

WO  stars represent a very advanced evolutionary stage of massive stars 
shortly before their supernova explosion. The WO winds are the fastest among 
all non-degenerate stars \citep{Sander2012,Tramper2015}.
Their wind opacity for  X-rays is 
less than that in WC stars (see Fig.\,\ref{fig:osktau1}), therefore X-rays may 
reach an observer.
\citet{osk2009} 
reported the first discovery of X-ray emission from this important type of 
objects. They detected the WO2 star WR142 using {\em XMM-Newton}. This X-ray 
source identification was confirmed with {\em Chandra}'s superb angular 
resolution \citep{Sokal2010}. 
The X-ray luminosity of WR142 is not very high  ($L_{\rm X}\approx 
10^{31}$\,erg\,s$^{-1}$), while its X-ray spectrum is quite hard (with 
$T_{\rm X}>100$\,MK). WO stars are basically bare stellar cores and might 
expose magnetic fields \citep{Fuller2015}.  This and other scenarios were 
considered in \citet{osk2009}, while \citet{Sokal2010} suggested the 
possibility 
of inverse Compton scattering or a binary companion.   

\section{Summary}

All types of single WR stars emit X-rays. Their X-ray luminosities are orders 
of magnitude lower than in persistent high-mass X-ray binaries. The X-ray 
spectra appear to  be thermal, with very hot plasma of a few$\times 10$\,MK 
being present along with cooler components. The mechanisms responsible for 
X-ray generation are not yet understood. Promising scenarios include an 
LDI-like mechanisms, interactions of wind streams with blobs, and CIRs. 
New observations and modeling shall uncover how X-rays are generated in
winds of single WR stars. 

\bibliographystyle{aa} 


\begin{thebibliography}{79}
\expandafter\ifx\csname natexlab\endcsname\relax\def\natexlab#1{#1}\fi

\bibitem[{{Babel} \& {Montmerle}(1997)}]{Babel1997}
{Babel}, J. \& {Montmerle}, T. 1997, \aap, 323, 121

\bibitem[{{Barnard} {et~al.}(2008){Barnard}, {Clark}, \& {Kolb}}]{Barnard2008}
{Barnard}, R., {Clark}, J.~S., \& {Kolb}, U.~C. 2008, \aap, 488, 697

\bibitem[{{Baum} {et~al.}(1992){Baum}, {Hamann}, {Koesterke}, \&
  {Wessolowski}}]{Baum1992}
{Baum}, E., {Hamann}, W.-R., {Koesterke}, L., \& {Wessolowski}, U. 1992, \aap,
  266, 402

\bibitem[{{Chen{\'e}} {et~al.}(2011){Chen{\'e}}, {Moffat}, {Cameron}, {Fahed},
  {Gamen}, {Lef{\`e}vre}, {Rowe}, {St-louis}, {Muntean}, {De La
  Chevroti{\`e}re}, {Guenther}, {Kuschnig}, {Matthews}, {Rucinski}, {Sasselov},
  \& {Weiss}}]{Chene2011}
{Chen{\'e}}, A.-N., {Moffat}, A.~F.~J., {Cameron}, C., {et~al.} 2011, \apj,
  735, 34

\bibitem[{{Chen{\'e}} \& {St-Louis}(2011)}]{Chenea2011}
{Chen{\'e}}, A.-N. \& {St-Louis}, N. 2011, \apj, 736, 140

\bibitem[{{Cherepashchuk}(1976)}]{Cherep1976}
{Cherepashchuk}, A.~M. 1976, Soviet Astronomy Letters, 2, 138

\bibitem[{{Clark} {et~al.}(2008){Clark}, {Muno}, {Negueruela}, {Dougherty},
  {Crowther}, {Goodwin}, \& {de Grijs}}]{Clark2008}
{Clark}, J.~S., {Muno}, M.~P., {Negueruela}, I., {et~al.} 2008, \aap, 477, 147

\bibitem[{{Cranmer} \& {Owocki}(1996)}]{Cranmer1996}
{Cranmer}, S.~R. \& {Owocki}, S.~P. 1996, \apj, 462, 469

\bibitem[{{Crowther} {et~al.}(1995){Crowther}, {Smith}, {Hillier}, \&
  {Schmutz}}]{Crow1995}
{Crowther}, P.~A., {Smith}, L.~J., {Hillier}, D.~J., \& {Schmutz}, W. 1995,
  \aap, 293, 427

\bibitem[{{de la Chevroti{\`e}re} {et~al.}(2014){de la Chevroti{\`e}re},
  {St-Louis}, {Moffat}, \& {MiMeS Collaboration}}]{Chev2014}
{de la Chevroti{\`e}re}, A., {St-Louis}, N., {Moffat}, A.~F.~J., \& {MiMeS
  Collaboration}. 2014, \apj, 781, 73

\bibitem[{{Feldmeier} {et~al.}(2003){Feldmeier}, {Oskinova}, \&
  {Hamann}}]{feld2003}
{Feldmeier}, A., {Oskinova}, L., \& {Hamann}, W.-R. 2003, \aap, 403, 217

\bibitem[{{Feldmeier} {et~al.}(1997){Feldmeier}, {Puls}, \&
  {Pauldrach}}]{feld1997}
{Feldmeier}, A., {Puls}, J., \& {Pauldrach}, A.~W.~A. 1997, \aap, 322, 878

\bibitem[{{Fuller} {et~al.}(2015){Fuller}, {Cantiello}, {Stello}, {Garcia}, \&
  {Bildsten}}]{Fuller2015}
{Fuller}, J., {Cantiello}, M., {Stello}, D., {Garcia}, R.~A., \& {Bildsten}, L.
  2015, ArXiv e-prints

\bibitem[{{Gayley}(2012)}]{Gayley2012}
{Gayley}, K.~G. 2012, in Astronomical Society of the Pacific Conference Series,
  Vol. 465, Proceedings of a Scientific Meeting in Honor of Anthony F. J.
  Moffat, ed. L.~{Drissen}, C.~{Robert}, N.~{St-Louis}, \& A.~F.~J. {Moffat},
  140

\bibitem[{{Gayley} \& {Owocki}(1995)}]{GO1995}
{Gayley}, K.~G. \& {Owocki}, S.~P. 1995, \apj, 446, 801

\bibitem[{{Gosset} {et~al.}(2005){Gosset}, {Naz{\'e}}, {Claeskens}, {Rauw},
  {Vreux}, \& {Sana}}]{Gosset2005}
{Gosset}, E., {Naz{\'e}}, Y., {Claeskens}, J.-F., {et~al.} 2005, \aap, 429, 685

\bibitem[{{Hamann} {et~al.}(2006){Hamann}, {Gr{\"a}fener}, \&
  {Liermann}}]{Hamann2006}
{Hamann}, W.-R., {Gr{\"a}fener}, G., \& {Liermann}, A. 2006, \aap, 457, 1015

\bibitem[{{Hyodo} {et~al.}(2008){Hyodo}, {Tsujimoto}, {Koyama}, {Nishiyama},
  {Nagata}, {Sakon}, {Murakami}, \& {Matsumoto}}]{Hyodo2008}
{Hyodo}, Y., {Tsujimoto}, M., {Koyama}, K., {et~al.} 2008, \pasj, 60, 173

\bibitem[{{Ignace}(2001)}]{Ignace2001}
{Ignace}, R. 2001, \apjl, 549, L119

\bibitem[{{Ignace} {et~al.}(2013){Ignace}, {Gayley}, {Hamann}, {Huenemoerder},
  {Oskinova}, {Pollock}, \& {McFall}}]{Ignace2013}
{Ignace}, R., {Gayley}, K.~G., {Hamann}, W.-R., {et~al.} 2013, \apj, 775, 29

\bibitem[{{Ignace} \& {Oskinova}(1999)}]{Ignace1999}
{Ignace}, R. \& {Oskinova}, L.~M. 1999, \aap, 348, L45

\bibitem[{{Ignace} {et~al.}(2003){Ignace}, {Oskinova}, \& {Brown}}]{Ignace2003}
{Ignace}, R., {Oskinova}, L.~M., \& {Brown}, J.~C. 2003, \aap, 408, 353

\bibitem[{{Ignace} {et~al.}(2000){Ignace}, {Oskinova}, \&
  {Foullon}}]{Ignace2000}
{Ignace}, R., {Oskinova}, L.~M., \& {Foullon}, C. 2000, \mnras, 318, 214

\bibitem[{{L{\'e}pine} \& {Moffat}(1999)}]{Lepine1999}
{L{\'e}pine}, S. \& {Moffat}, A.~F.~J. 1999, \apj, 514, 909

\bibitem[{{Leutenegger} {et~al.}(2006){Leutenegger}, {Paerels}, {Kahn}, \&
  {Cohen}}]{Leu2006}
{Leutenegger}, M.~A., {Paerels}, F.~B.~S., {Kahn}, S.~M., \& {Cohen}, D.~H.
  2006, \apj, 650, 1096

\bibitem[{{Lobel} \& {Blomme}(2008)}]{Lobel2008}
{Lobel}, A. \& {Blomme}, R. 2008, \apj, 678, 408

\bibitem[{{Lomax} {et~al.}(2015){Lomax}, {Naz{\'e}}, {Hoffman}, {Russell}, {De
  Becker}, {Corcoran}, {Davidson}, {Neilson}, {Owocki}, {Pittard}, \&
  {Pollock}}]{Lomax2015}
{Lomax}, J.~R., {Naz{\'e}}, Y., {Hoffman}, J.~L., {et~al.} 2015, \aap, 573, A43

\bibitem[{{Macfarlane} {et~al.}(1991){Macfarlane}, {Cassinelli}, {Welsh},
  {Vedder}, {Vallerga}, \& {Waldron}}]{Mac1991}
{Macfarlane}, J.~J., {Cassinelli}, J.~P., {Welsh}, B.~Y., {et~al.} 1991, \apj,
  380, 564

\bibitem[{{Massa} {et~al.}(2014){Massa}, {Oskinova}, {Fullerton}, {Prinja},
  {Bohlender}, {Morrison}, {Blake}, \& {Pych}}]{massa2014}
{Massa}, D., {Oskinova}, L., {Fullerton}, A.~W., {et~al.} 2014, \mnras, 441,
  2173

\bibitem[{{Mauerhan} {et~al.}(2011){Mauerhan}, {Van Dyk}, \&
  {Morris}}]{Mau2011}
{Mauerhan}, J.~C., {Van Dyk}, S.~D., \& {Morris}, P.~W. 2011, \aj, 142, 40

\bibitem[{{Michaux} {et~al.}(2014){Michaux}, {Moffat}, {Chen{\'e}}, \&
  {St-Louis}}]{Mich2014}
{Michaux}, Y.~J.~L., {Moffat}, A.~F.~J., {Chen{\'e}}, A.-N., \& {St-Louis}, N.
  2014, \mnras, 440, 2

\bibitem[{{Morel} {et~al.}(1997){Morel}, {St-Louis}, \&
  {Marchenko}}]{Morel1997}
{Morel}, T., {St-Louis}, N., \& {Marchenko}, S.~V. 1997, \apj, 482, 470

\bibitem[{{Mullan}(1984)}]{Mullan1984}
{Mullan}, D.~J. 1984, \apj, 283, 303

\bibitem[{{Naz{\'e}}(2009)}]{naze2009}
{Naz{\'e}}, Y. 2009, \aap, 506, 1055

\bibitem[{{Naz{\'e}} {et~al.}(2013){Naz{\'e}}, {Oskinova}, \&
  {Gosset}}]{naze2013}
{Naz{\'e}}, Y., {Oskinova}, L.~M., \& {Gosset}, E. 2013, \apj, 763, 143

\bibitem[{{Nebot G{\'o}mez-Mor{\'a}n} {et~al.}(2015){Nebot
  G{\'o}mez-Mor{\'a}n}, {Motch}, {Pineau}, {Carrera}, {Pakull}, \&
  {Riddick}}]{nebot2015}
{Nebot G{\'o}mez-Mor{\'a}n}, A., {Motch}, C., {Pineau}, F.-X., {et~al.} 2015,
  \mnras, 452, 884

\bibitem[{{Oskinova}(2005)}]{osk2005}
{Oskinova}, L.~M. 2005, \mnras, 361, 679

\bibitem[{{Oskinova} {et~al.}(2001){Oskinova}, {Clarke}, \&
  {Pollock}}]{osk2001}
{Oskinova}, L.~M., {Clarke}, D., \& {Pollock}, A.~M.~T. 2001, \aap, 378, L21

\bibitem[{{Oskinova} {et~al.}(2004){Oskinova}, {Feldmeier}, \&
  {Hamann}}]{osk2004}
{Oskinova}, L.~M., {Feldmeier}, A., \& {Hamann}, W.-R. 2004, \aap, 422, 675

\bibitem[{{Oskinova} {et~al.}(2012){Oskinova}, {Gayley}, {Hamann},
  {Huenemoerder}, {Ignace}, \& {Pollock}}]{osk2012}
{Oskinova}, L.~M., {Gayley}, K.~G., {Hamann}, W.-R., {et~al.} 2012, \apjl, 747,
  L25

\bibitem[{{Oskinova} \& {Hamann}(2008)}]{osk2008}
{Oskinova}, L.~M. \& {Hamann}, W.-R. 2008, \mnras, 390, L78

\bibitem[{{Oskinova} {et~al.}(2011){Oskinova}, {Hamann}, {Cassinelli}, {Brown},
  \& {Todt}}]{osk2011}
{Oskinova}, L.~M., {Hamann}, W.-R., {Cassinelli}, J.~P., {Brown}, J.~C., \&
  {Todt}, H. 2011, Astronomische Nachrichten, 332, 988

\bibitem[{{Oskinova} {et~al.}(2009){Oskinova}, {Hamann}, {Feldmeier}, {Ignace},
  \& {Chu}}]{osk2009}
{Oskinova}, L.~M., {Hamann}, W.-R., {Feldmeier}, A., {Ignace}, R., \& {Chu},
  Y.-H. 2009, \apjl, 693, L44

\bibitem[{{Oskinova} {et~al.}(2003){Oskinova}, {Ignace}, {Hamann}, {Pollock},
  \& {Brown}}]{osk2003}
{Oskinova}, L.~M., {Ignace}, R., {Hamann}, W.-R., {Pollock}, A.~M.~T., \&
  {Brown}, J.~C. 2003, \aap, 402, 755

\bibitem[{{Owocki} {et~al.}(2013){Owocki}, {Sundqvist}, {Cohen}, \&
  {Gayley}}]{owocki2013}
{Owocki}, S.~P., {Sundqvist}, J.~O., {Cohen}, D.~H., \& {Gayley}, K.~G. 2013,
  \mnras, 429, 3379

\bibitem[{{Pallavicini} {et~al.}(1981){Pallavicini}, {Golub}, {Rosner},
  {Vaiana}, {Ayres}, \& {Linsky}}]{Pallavicini1981}
{Pallavicini}, R., {Golub}, L., {Rosner}, R., {et~al.} 1981, \apj, 248, 279

\bibitem[{{Pandey} {et~al.}(2014){Pandey}, {Pandey}, \&
  {Karmakar}}]{Pandey2014}
{Pandey}, J.~C., {Pandey}, S.~B., \& {Karmakar}, S. 2014, \apj, 788, 84

\bibitem[{{Parkin} \& {Sim}(2013)}]{Parkin2013}
{Parkin}, E.~R. \& {Sim}, S.~A. 2013, \apj, 767, 114

\bibitem[{{Pittard} \& {Dougherty}(2006)}]{Pittard2006}
{Pittard}, J.~M. \& {Dougherty}, S.~M. 2006, \mnras, 372, 801

\bibitem[{{Pollock}(1987{\natexlab{a}})}]{Pollockb1987}
{Pollock}, A.~M.~T. 1987{\natexlab{a}}, \aap, 171, 135

\bibitem[{{Pollock}(1987{\natexlab{b}})}]{Pollock1987}
{Pollock}, A.~M.~T. 1987{\natexlab{b}}, \apj, 320, 283

\bibitem[{{Pollock} {et~al.}(2005){Pollock}, {Corcoran}, {Stevens}, \&
  {Williams}}]{Pollock2005}
{Pollock}, A.~M.~T., {Corcoran}, M.~F., {Stevens}, I.~R., \& {Williams}, P.~M.
  2005, \apj, 629, 482

\bibitem[{{Pollock} {et~al.}(1995){Pollock}, {Haberl}, \&
  {Corcoran}}]{Pollock1995}
{Pollock}, A.~M.~T., {Haberl}, F., \& {Corcoran}, M.~F. 1995, in IAU Symposium,
  Vol. 163, Wolf-Rayet Stars: Binaries; Colliding Winds; Evolution, ed. K.~A.
  {van der Hucht} \& P.~M. {Williams}, 512

\bibitem[{{Rauw} \& {Naze}(2015)}]{Rauw2015}
{Rauw}, G. \& {Naze}, Y. 2015, ArXiv e-prints

\bibitem[{{Rauw} {et~al.}(2015){Rauw}, {Naz{\'e}}, {Wright}, {Drake},
  {Guarcello}, {Prinja}, {Peck}, {Albacete Colombo}, {Herrero}, {Kobulnicky},
  {Sciortino}, \& {Vink}}]{Rauwa2015}
{Rauw}, G., {Naz{\'e}}, Y., {Wright}, N.~J., {et~al.} 2015, \apjs, 221, 1

\bibitem[{{Sander} {et~al.}(2012){Sander}, {Hamann}, \& {Todt}}]{Sander2012}
{Sander}, A., {Hamann}, W.-R., \& {Todt}, H. 2012, \aap, 540, A144

\bibitem[{{Sanders} {et~al.}(1985){Sanders}, {Cassinelli}, {Myers}, \& {van der
  Hucht}}]{Sanders1985}
{Sanders}, W.~T., {Cassinelli}, J.~P., {Myers}, R.~V., \& {van der Hucht},
  K.~A. 1985, \apj, 288, 756

\bibitem[{{Schild} {et~al.}(2004){Schild}, {G{\"u}del}, {Mewe}, {Schmutz},
  {Raassen}, {Audard}, {Dumm}, {van der Hucht}, {Leutenegger}, \&
  {Skinner}}]{Schild2004}
{Schild}, H., {G{\"u}del}, M., {Mewe}, R., {et~al.} 2004, \aap, 422, 177

\bibitem[{{Seward} {et~al.}(1979){Seward}, {Forman}, {Giacconi}, {Griffiths},
  {Harnden}, {Jones}, \& {Pye}}]{Seward1979}
{Seward}, F.~D., {Forman}, W.~R., {Giacconi}, R., {et~al.} 1979, \apjl, 234,
  L55

\bibitem[{{Shaviv}(2000)}]{Shaviv2000}
{Shaviv}, N.~J. 2000, \apjl, 532, L137

\bibitem[{{Shenar} {et~al.}(2014){Shenar}, {Hamann}, \& {Todt}}]{Shenar2014}
{Shenar}, T., {Hamann}, W.-R., \& {Todt}, H. 2014, \aap, 562, A118

\bibitem[{{Skinner} {et~al.}(2002{\natexlab{a}}){Skinner}, {Zhekov},
  {G{\"u}del}, \& {Schmutz}}]{Skina2002}
{Skinner}, S.~L., {Zhekov}, S.~A., {G{\"u}del}, M., \& {Schmutz}, W.
  2002{\natexlab{a}}, \apj, 579, 764

\bibitem[{{Skinner} {et~al.}(2002{\natexlab{b}}){Skinner}, {Zhekov},
  {G{\"u}del}, \& {Schmutz}}]{Skinb2002}
{Skinner}, S.~L., {Zhekov}, S.~A., {G{\"u}del}, M., \& {Schmutz}, W.
  2002{\natexlab{b}}, \apj, 572, 477

\bibitem[{{Skinner} {et~al.}(2010){Skinner}, {Zhekov}, {G{\"u}del}, {Schmutz},
  \& {Sokal}}]{Skin2010}
{Skinner}, S.~L., {Zhekov}, S.~A., {G{\"u}del}, M., {Schmutz}, W., \& {Sokal},
  K.~R. 2010, \aj, 139, 825

\bibitem[{{Skinner} {et~al.}(2012){Skinner}, {Zhekov}, {G{\"u}del}, {Schmutz},
  \& {Sokal}}]{Skin2012}
{Skinner}, S.~L., {Zhekov}, S.~A., {G{\"u}del}, M., {Schmutz}, W., \& {Sokal},
  K.~R. 2012, \aj, 143, 116

\bibitem[{{Sokal} {et~al.}(2010){Sokal}, {Skinner}, {Zhekov}, {G{\"u}del}, \&
  {Schmutz}}]{Sokal2010}
{Sokal}, K.~R., {Skinner}, S.~L., {Zhekov}, S.~A., {G{\"u}del}, M., \&
  {Schmutz}, W. 2010, \apj, 715, 1327

\bibitem[{{Stevens} {et~al.}(1992){Stevens}, {Blondin}, \&
  {Pollock}}]{Stev1992}
{Stevens}, I.~R., {Blondin}, J.~M., \& {Pollock}, A.~M.~T. 1992, \apj, 386, 265

\bibitem[{{Stevens} {et~al.}(1996){Stevens}, {Corcoran}, {Willis}, {Skinner},
  {Pollock}, {Nagase}, \& {Koyama}}]{Stevens1996}
{Stevens}, I.~R., {Corcoran}, M.~F., {Willis}, A.~J., {et~al.} 1996, \mnras,
  283, 589

\bibitem[{{Tramper} {et~al.}(2015){Tramper}, {Straal}, {Sanyal}, {Sana}, {de
  Koter}, {Gr{\"a}fener}, {Langer}, {Vink}, {de Mink}, \&
  {Kaper}}]{Tramper2015}
{Tramper}, F., {Straal}, S.~M., {Sanyal}, D., {et~al.} 2015, \aap, 581, A110

\bibitem[{{Waldron} \& {Cassinelli}(2007)}]{WC2007}
{Waldron}, W.~L. \& {Cassinelli}, J.~P. 2007, \apj, 668, 456

\bibitem[{{Wessolowski}(1996)}]{Wes1996}
{Wessolowski}, U. 1996, in Roentgenstrahlung from the Universe, ed. H.~U.
  {Zimmermann}, J.~{Tr{\"u}mper}, \& H.~{Yorke}, 75--76

\bibitem[{{White} \& {Long}(1986)}]{WL1986}
{White}, R.~L. \& {Long}, K.~S. 1986, \apj, 310, 832

\bibitem[{{Williams} {et~al.}(1987){Williams}, {van der Hucht}, \&
  {The}}]{Wil1987}
{Williams}, P.~M., {van der Hucht}, K.~A., \& {The}, P.~S. 1987, \qjras, 28,
  248

\bibitem[{{Willis} \& {Stevens}(1996)}]{Willis1996}
{Willis}, A.~J. \& {Stevens}, I.~R. 1996, \aap, 310, 577

\bibitem[{{Wrigge} {et~al.}(1994){Wrigge}, {Wendker}, \&
  {Wisotzki}}]{Wrigge1994}
{Wrigge}, M., {Wendker}, H.~J., \& {Wisotzki}, L. 1994, \aap, 286, 219

\bibitem[{{Zhekov}(2007)}]{Zhekov2007}
{Zhekov}, S.~A. 2007, \mnras, 382, 886

\bibitem[{{Zhekov}(2012)}]{Zhekov2012}
{Zhekov}, S.~A. 2012, \mnras, 422, 1332

\bibitem[{{Zhekov} {et~al.}(2011){Zhekov}, {Gagn{\'e}}, \&
  {Skinner}}]{Zhekov2011}
{Zhekov}, S.~A., {Gagn{\'e}}, M., \& {Skinner}, S.~L. 2011, \apjl, 727, L17

\bibitem[{{Zhekov} \& {Skinner}(2015)}]{Zhekov2015}
{Zhekov}, S.~A. \& {Skinner}, S.~L. 2015, \mnras, 452, 872

\end{thebibliography}

\end{document}